\documentclass[twocolumn]{aastex62}

\usepackage{graphicx}
\usepackage{amsmath}
\usepackage{float}
\usepackage{natbib}

\graphicspath{{./}{figures/}}

\shorttitle{MSPs and BHs in GCs}
\shortauthors{Ye et al.}

\begin{document}

\title{Millisecond Pulsars and Black Holes in Globular Clusters}

\author[0000-0001-9582-881X]{Claire S.\ Ye}
\affil{ Department of Physics \& Astronomy, Northwestern University, Evanston, IL 60208, USA}
\affil{ Center for Interdisciplinary Exploration \& Research in Astrophysics (CIERA), Evanston, IL 60208, USA}
\correspondingauthor{Claire S.\ Ye}
\email{shiye2015@u.northwestern.edu}

\author[0000-0002-4086-3180]{Kyle Kremer}
\affil{ Department of Physics \& Astronomy, Northwestern University, Evanston, IL 60208, USA}
\affil{ Center for Interdisciplinary Exploration \& Research in Astrophysics (CIERA), Evanston, IL 60208, USA}

\author[0000-0002-3680-2684]{Sourav Chatterjee}
\affil{Tata Institute of Fundamental Research, Homi Bhabha Road, Mumbai 400005, India}
\affil{ Center for Interdisciplinary Exploration \& Research in Astrophysics (CIERA), Evanston, IL 60208, USA}

\author{Carl L.\ Rodriguez}
\affil{MIT-Kavli Institute for Astrophysics and Space Research, Cambridge, MA 02139, USA}

\author[0000-0002-7132-418X]{Frederic A.\ Rasio}
\affil{ Department of Physics \& Astronomy, Northwestern University, Evanston, IL 60208, USA}
\affil{ Center for Interdisciplinary Exploration \& Research in Astrophysics (CIERA), Evanston, IL 60208, USA}

\begin{abstract}
Over a hundred millisecond radio pulsars (MSPs) have been observed in globular clusters (GCs), motivating theoretical studies of the formation and evolution of these sources through stellar evolution coupled to stellar dynamics. Here we study MSPs in GCs using realistic $N$-body simulations with our Cluster Monte Carlo code. We show that neutron stars (NSs) formed in electron-capture supernovae (including both accretion-induced and merger-induced collapse of white dwarfs) can be spun up through mass transfer to form MSPs. Both NS formation and spin-up through accretion are greatly enhanced through dynamical interaction processes. We find that our models for average GCs at the present day with masses $\approx 2 \times 10^5\,M_\odot$ can produce up to $10-20$ MSPs, while a very massive GC model with mass $\approx 10^6\,M_\odot$ can produce close to $100$. We show that the number of MSPs is anti-correlated with the total number of stellar-mass black holes (BHs) retained in the host cluster. The radial distributions are also affected: MSPs are more concentrated towards the center in a host cluster with a smaller number of retained BHs. As a result, the number of MSPs in a GC could be used to place constraints on its BH population. Some intrinsic properties of MSP systems in our models  (such as the magnetic fields and spin periods) are in good overall agreement with observations, while others (such as the distribution of binary companion types) less so, and we discuss the possible reasons for such discrepancies. Interestingly, our models also demonstrate the possibility of dynamically forming NS--NS and NS--BH binaries in GCs, although the predicted numbers are very small.
\end{abstract}


\section{Introduction} \label{sec:intro}
Globular clusters (GCs) are known to be highly efficient at producing millisecond pulsars (MSPs). Since the discovery of radio MSPs in GCs in the 1980s \citep{lyne1987discovery}, multiple pulsar surveys have found 150 pulsars in 28 GCs\footnote{GC pulsar catalog: \url{http://www.naic.edu/~pfreire/GCpsr.html}}  \citep[for reviews, see][]{camilo2005pulsars,ransom2008}, including 38 in Terzan~5 and 25 in 47~Tuc. Although GCs make up only about $0.05\%$
of stars in the Milky Way, collectively, GCs contain more than one third of the total number of known MSPs in our Galaxy \citep{manchester2005australia}\footnote{ATNF pulsar catalog: \url{http://www.atnf.csiro.au/research/pulsar/psrcat/}}.

GCs also contain many low-mass X-ray binaries \citep[LMXBs;][]{clark1975x} with neutron star (NS) accretors. The low surface magnetic fields ($\sim 10^7 - 10^9\,$G) and short spin periods ($\lesssim 30\,$ms) of MSPs suggest that they are ``recycled" pulsars \citep{alpar1982new} with LMXBs as their likely progenitors. Indeed some ``transitional'' MSPs, providing the link between LMXBs and  MSPs, have recently been detected; these are observed to switch back and forth between phases of accretion-powered X-ray emission and rotation-powered radio emission. At present, there are three confirmed transitional MSPs, including one in a GC, M28 \citep{papitto2013swings}, and two in the Galactic field \citep{archibald2009radio,bassa2014state,roy2015discovery}. There are also a few additional candidates in GCs and in the field \citep[][and references therein]{bahramian2018maveric}.
While the physics is far from being understood in detail, it is plausible that mass transfer onto old, slowly spinning NSs can ``bury" their magnetic fields while at the same time spinning them up \citep[e.g.][and references therein]{bhattacharya1991formation,rappaport1995relation,Kiel_2008,tauris2012formation}. In addition, MSPs have low spin-down rates and thus long lifetimes ($\sim 10^{10}\,\rm{yr}$, compared to $\sim 10^7\,$yr for young pulsars), which makes them easier to observe in old stellar systems like Milky Way GCs.

The large numbers of MSPs and NS LMXBs suggest that a typical Galactic GC on average contain at least a few hundred NSs \citep[e.g.,][]{Ivanova_2008}. However, observations show that the majority of NSs in the Galactic field are born with velocities $\gtrsim 200\,\rm{km\,s^{-1}}$ due to natal kicks associated with asymmetries in core collapse supernovae (CCSNe); see, e.g., \citet{hobbs2005statistical}. Because of these large natal kicks, the majority of CCSN NSs born in GCs (where escape velocities are generally $<50-100\,\rm{km\,s^{-1}}$, even at early times when the clusters may have been more massive than at present) are ejected from the cluster at birth. This is seemingly at odds with the large numbers of NSs inferred to be present in GCs, which is often referred to as the NS ``retention problem'' \citep[e.g.][]{Pfahl_2002}. However, the discovery of high-mass X-ray binaries with long orbital periods ($P_{orb}>30\,\rm{days}$) and low eccentricities ($e\lesssim0.2$) \citep[HMXBs;][]{pfahl2002new} suggests that some NSs must be born with very small natal kicks. Additionally, NSs born in massive binaries may be easier to retain in GCs, but \cite{Pfahl_2002} showed that the retention fraction of NSs formed through CCSNe in massive binaries is still at most a few percent, not enough to explain the large populations of NSs in GCs. 

Later studies suggested that electron-capture supernovae (ECSNe) can solve the retention problem by producing many NSs with small natal kicks \citep{Podsiadlowski_2004,Ivanova_2008}. When electron capture occurs onto $\rm{Mg}^{24}$ and $\rm{Ne}^{20}$ in a degenerate ONeMg core, it triggers the core to collapse to a NS \citep{miyaji1980supernova,nomoto1984evolution,nomoto1987evolution}. The explosion energy of ECSNe is much lower than that of CCSNe for Fe cores. As a result, NSs formed in ECSNe could receive an order of magnitude smaller natal kicks compared to NSs formed in CCSNe \citep{Podsiadlowski_2004}. Therefore a large fraction of ECSN NSs can be retained in GCs after formation \citep{kuranov2006neutron,Ivanova_2008}, in contrast to CCSN NSs, which are mostly lost.

The high stellar densities in GC cores lead to frequent dynamical encounters and high formation rates of NSs, MSPs and LMXBS \citep{clark1975x,hut1992binaries,pooley2003dynamical,hui2010dynamical,bahramian2013stellar}. For example, dynamical interactions can enhance the WD--WD merger rate \citep[e.g.][]{shara2002star}, which can lead to a higher merger-induced collapse rate of WDs to NSs \citep{1985A&A...150L..21S}. NS binaries are also created at an increased rate since single NSs can acquire companions through exchange interactions \citep{sigurdsson1993,Sigurdsson1995binaryandneutronstar,rasio2000formation,Ivanova_2008}. Subsequent stellar evolution of the companion and
hardening through repeated close flybys can then trigger Roche-lobe overflow (RLOF) and mass transfer, possibly turning the NS into a MSP. More than half of the observed MSPs in GCs are known to be in binaries.

Several studies have shown that the number of MSPs in GCs is correlated with the stellar encounter rate, $\propto \rho_0^2r_c^3/\sigma_0$, where $\rho_0$ is the central luminosity density, $r_c$ is the core radius and $\sigma_0$ is the central velocity dispersion \citep{Verbunt1987xraybinary,hui2010dynamical,bahramian2013stellar}. In clusters with larger encounter rates, NSs undergo more dynamical interactions, resulting in more NSs in binaries and binaries with shorter orbital periods. For example, while double NSs (DNSs) are very rare, the only confirmed GC DNS is in M15, which is a core-collapsed cluster with extremely high central density. It has been suggested that the frequent stellar encounters in M15 led to the formation of this DNS \citep{anderson1990discovery,prince1991timing,Phinney:1991aa,deich1996massm15c,jacoby2006measurement}. Additionally, BHs could have a strong influence on the formation of NS binaries by altering the evolution of their host GCs. Once formed, BHs quickly mass-segregate to the cluster core through dynamical friction. Several recent studies have shown that the long-term retention fraction of BHs in GCs strongly affects the GC core densities and the ability of other compact objects to participate in the core dynamics \citep{Mackey_2008,Morscher_2015,Chatterjee_2017,arca2018mocca,Kremer_2018,fragione2018neutron}.

Very few previous works have attempted to study the formation and evolution of MSPs in the context of fully realistic $N$-body simulations of GCs. \cite{Ivanova_2008} modeled the formation and evolution of NSs in GCs using the population synthesis code {\tt StarTrack} \citep{belczynski2002comprehensive,belczynski2008compact} to follow single and binary star evolution in a fixed cluster background, and the small $N$-body integrator {\tt Fewbody} \citep{fregeau2004stellar} to compute dynamical interactions in the cluster core. They included ECSNe for NS formation and showed that the low natal kicks associated with these NSs were crucial for matching the observed numbers of MSPs in GCs. Here, we build upon this previous study by performing full, self-consistent $N$-body simulations for the cluster dynamics, and also incorporating the effects of NS magnetic field and spin period evolution during mass-transfer.

In Section~\ref{sec:method}, we describe how we model MSPs in GCs. We describe the set of models used for this study and our methods and assumptions for tracking the magnetic fields and spin periods of NSs. We also provide some examples of the complicated dynamical evolution of NSs in our models. In Section~\ref{sec:results} we provide an overview of results. We explore the expected anti-correlation between the numbers of MSPs and BHs in GCs. We also compare our results broadly with pulsar observations. We discuss selection effects and summarize our findings in Section~\ref{sec:sum_discuss}.

\section{METHODS} \label{sec:method}
\subsection{Models}\label{subsec:models}
We use our \texttt{Cluster Monte Carlo} code (\texttt{CMC}) to simulate a grid of models for this study. \texttt{CMC} is a parallelized H\'{e}non-type Monte Carlo code \citep{henon1971monte,henon1971montecluster} which has been developed and rigorously tested over many years \citep{Joshi_2000,Joshi_2001,Fregeau_2003,fregeau2007monte, Chatterjee_2010,Umbreit_2012,Pattabiraman_2013,Chatterjee_2013a,rodriguez2018post}. \texttt{CMC} incorporates all the relevant physics for GC evolution, including two-body relaxation, three-body binary formation, strong three- and four-body interactions, and some post-Newtonian effects \citep{rodriguez2018post}. Updated versions of the \texttt{SSE} and \texttt{BSE} packages \citep{hurley2000comprehensive,hurley2002evolution} are used for single and binary stellar evolution in \texttt{CMC} and \texttt{Fewbody} \citep{fregeau2004stellar,fregeau2007monte} is used to directly integrate all three- and four-body gravitational encounters, with post-Newtonian effects \citep{antognini2014rapid,amaro2016relativistic}.

In this study, we consider a set of 26 independent cluster models. In models 1-25 (which serve as our main grid of models with present-day properties typical of Milky Way GCs), we fix a number of initial cluster parameters: total star number $N=8 \times 10^5$, binary fraction $f_b=5\%$, virial radius $r_v=1\,\rm{pc}$, King concentration parameter $W_o=5$, galactocentric distance $r_g=8\,\rm{kpc}$ and metallicity $\rm{Z} = 0.001$ \citep{Morscher_2015,Chatterjee_2017}. We use the initial mass function given in \cite{kroupa2001variation} ranging from $0.08$ to $150\,M_{\odot}$ to sample the initial stellar masses. NS remnants from CCSNe at formation receive natal kicks drawn from a Maxwellian distribution with a standard deviation $\sigma_{NS}=265\,\rm{km\,s^{-1}}$ \citep{hobbs2005statistical}.

Following \citet{Kremer_2018}, the only parameter varied in our models is the natal kick for BHs, which allows us to easily isolate and understand the effects of BH retention on the long-term GC evolution and the dynamical evolution of MSPs. Varying the BH natal kicks to alter the retained number of BHs at present times has a similar effect on the GC properties as varying more physically motivated initial conditions, such as the initial virial radius \citep{kremer2018initial}. With all other parameters being fixed, {\em all} differences between models clearly originate from the differences in the fraction of BHs retained in the cluster upon formation \citep[for more details see][]{Kremer_2018}. All models were evolved for $12\,\rm{Gyr}$.

The natal kicks for BHs are assumed to be independent of BH masses and are first drawn from the same Maxwellian distribution as the NSs. Their magnitudes are then multiplied by the ratio \(\frac{\sigma_{BH}}{\sigma_{NS}}\) which is varied between models. We set different ratios from 0.005 to 1.0 for different models as shown in Table~\ref{tab:table_1}. The introduction of this simple parameter controlling the natal kicks received by the BHs gives us a key benefit for the purpose of this initial study: it allows us to control the BH retention fractions of our models without the necessity to change any other initial parameters. As a result, the changes between the properties of these models can unambiguously be attributed to the difference in BH retention fractions.

In addition, we simulate one model (model 26 in Table~\ref{tab:table_1}) meant to represent a very massive GC. This model has initial $N=3.5\times10^6$, binary fraction $10\%$ and BH natal kick $\frac{\sigma_{BH}}{\sigma_{NS}}=1.0$. Other initial conditions for this model are the same as in models 1-25. The final mass of the model is about $1.2 \times 10^6\,M_{\odot}$, which is close to the masses of 47~Tuc and Terzan~5. This model is an extreme case designed to form many MSPs as a result of the large initial $N$, higher binary fraction, and large BH kicks.

Throughout this paper, we refer to models with final numbers of retained BHs greater than 200 as ``BH-rich" models, while those with $< 10$ BHs at $12\,$Gyr are called ``BH-poor" models; others in between these limits are referred to as ``BH-intermediate" models.

\startlongtable
\begin{deluxetable*}{ccc|ccccccccccccc}
\tabletypesize{\scriptsize}
\tablewidth{0pt}
\tablecaption{Cluster Model Properties \label{tab:table_1}}
\tablehead{
\colhead{$Model$} & \colhead{$N$} & \colhead{$\frac{\sigma_{BH}}{\sigma_{NS}}$} & \colhead{$r_c$} & \colhead{$r_{hl}$} & \colhead{$M_{TOT}$} & \colhead{$N_{BH}$} & \colhead{$N_{NS}$} & \colhead{$N_{NS-DYN}$} & \colhead{$N_{PSR}$} & \colhead{$N_{MSP}$} & \colhead{$N_{sMSP}$} & \colhead{$N_{bMSP}$} & \colhead{$NS-BH$} & \colhead{$DNS$}\\
\colhead{} &  $10^5$ & \colhead{} & \multicolumn{2}{c}{$\rm{pc}$} & \colhead{$10^5\,M_{\odot}$} & \colhead{} & \colhead{} & \colhead{} & \colhead{} & \colhead{} & \colhead{} & \colhead{} & \multicolumn{2}{c}{\rm{$9\,\rm{Gyr}<t<12\,\rm{Gyr}$}}
}
\startdata
1 & 8 & 0.005 & 2.88 & 4.00 & 2.10 & 383 & 444 & 15 & 1 & 1 & 0 & 1 & 0 & 0 \\
2 & 8 & 0.01 & 1.85 & 4.45 & 2.12 & 366 & 405 & 12 & 2 & 2 & 1 & 1  & 0 & 0 \\
3 & 8 & 0.02 & 2.72 & 4.04 & 2.12 & 332 & 414 & 20 & 1 & 1 & 0 & 1  & 0 & 0 \\
4 & 8 & 0.03 & 1.85 & 3.92 & 2.14 & 339 & 431 & 14 & 1 & 1 & 0 & 1  & 0 & 0 \\
5 & 8 & 0.04 & 1.54 & 4.17 & 2.15 & 328 & 396 & 11 & 0 & 0 & 0 & 0  & 0 & 0 \\
6 & 8 & 0.05 & 2.31 & 4.39 & 2.11 & 338 & 424 & 20 & 2 & 2 & 0 & 2  & 0 & 0 \\
7 & 8 & 0.06 & 2.04 & 3.49 & 2.15 & 293 & 446 & 14 & 1 & 1 & 0 & 1  & 0 & 0 \\
8 & 8 & 0.07 & 1.53 & 3.73 & 2.18 & 277 & 437 & 12 & 1 & 1 & 0 & 1  & 0 & 0 \\
9 & 8 & 0.08 & 1.44 & 3.18 & 2.20 & 237 & 414 & 14 & 1 & 1 & 0 & 1  & 0 & 0 \\
10 & 8 & 0.09 & 1.53 & 3.10 & 2.22 & 210 & 431 & 13 & 1 & 1 & 0 & 1 & 0 & 0 \\
11 & 8 & 0.1 &  1.20 & 3.02 & 2.24 & 189 & 489 & 21 & 1 & 1 & 0 & 1 & 1 & 0 \\
12 & 8 & 0.11 & 0.68 & 2.60 & 2.28 & 136 & 445 & 18 & 2 & 2 & 0 & 2 & 0 & 0 \\
13 & 8 & 0.12 & 0.77 & 2.31 & 2.29 & 114 & 452 & 20 & 1 & 0 & 0 & 0 & 0 & 0 \\
14 & 8 & 0.13 & 0.51 & 1.79 & 2.31 & 90 & 440 & 21 & 1 & 1 & 0 & 1  & 0 & 0 \\
15 & 8 & 0.14 & 0.54 & 1.83 & 2.31 & 66 & 464 & 18 & 0 & 0 & 0 & 0  & 0 & 0 \\
16 & 8 & 0.15 & 0.48 & 1.82 & 2.32 & 67 & 473 & 20 & 1 & 1 & 0 & 1  & 0 & 0 \\
17 & 8 & 0.16 & 0.18 & 1.53 & 2.33 & 39 & 462 & 14 & 2 & 2 & 1 & 1  & 2 & 0 \\
18 & 8 & 0.17 & 0.36 & 1.44 & 2.35 & 31 & 436 & 19 & 5 & 4 & 0 & 4  & 0 & 0 \\
19 & 8 & 0.18 & 0.24 & 1.51 & 2.35 & 21 & 443 & 27 & 3 & 1 & 0 & 1  & 0 & 0 \\
20 & 8 & 0.19 & 0.14 & 1.43 & 2.29 & 11 & 453 & 32 & 1 & 0 & 0 & 0  & 1 & 0 \\
21 & 8 & 0.2 & 0.25 & 1.48 & 2.32 & 5 & 493 & 48 & 4 & 1 & 0 & 1    & 3 & 2\\
22 & 8 & 0.4 & 0.20 & 1.95 & 2.29 & 1 & 505 & 178 & 8 & 7 & 1 & 6   & 2 & 10\\
23 & 8 & 0.6 & 0.34 & 2.07 & 2.34 & 1 & 481 & 173 & 9 & 8 & 3 & 5   & 0 & 7 \\
24 & 8 & 0.8 & 0.22 & 2.05 & 2.31 & 1 & 522 & 187 & 17 & 13 & 1 & 12 & 7 & 15 \\
25 & 8 & 1.0 & 0.24 & 2.03 & 2.28 & 0 & 524 & 184 & 10 & 7 & 2 & 5  & 0 & 4 \\
26 & 35 & 1.0 & 0.18 & 1.41 & 11.18 & 20 & 3908 & 1211 & 132 & 83 & 14 & 69 & 24 & 55 \\
\enddata
\tablecomments{Column 1--3: model number, initial number of stars and BH natal kick scaling factor, respectively. Columns 4--13 show final model properties at $12\,\rm{Gyr}$: projected core radius, projected half-light radius, total cluster mass, number of BHs, number of NSs, number of dynamically-formed NSs, total number of pulsars including MSPs, total number of MSPs, number of single MSPs, and number of binary MSPs. The last two columns give the number of NS--BH binaries and DNSs that appear at any time in the model between $9-12\,\rm{Gyr}$. $N_{NS-DYN}$ is the number of NSs formed, or strongly affected, by dynamical processes, and includes all NSs formed through MIC or AIC, as well as all NSs that experienced a direct collision or merger with another star and had their properties reset (as explained in Sec.~\ref{subsubsec:BP})}
\end{deluxetable*}

\subsection{Simulating NS evolution}\label{subsec:simuNS}
For this work we have updated the prescriptions for the evolution of NSs from \texttt{SSE} and \texttt{BSE} in \texttt{CMC}. The version of \texttt{BSE} used in \texttt{CMC} is now nearly identical to \texttt{COSMIC 2.0.0} \citep{breivik_katelyn_2019_2642803}. These updates include changes to the magnetic field and spin-period evolution for single and binary NSs, and to the natal kick prescriptions for NSs formed in ECSNe \citep{Kiel_2008,kiel2009populating}.

\subsubsection{Magnetic Field and Spin-Period Evolution}\label{subsubsec:BP}
The evolution of NS magnetic fields and spin periods has long been a topic of debate and still remains uncertain \citep[e.g.,][and references therein]{faucher2006birth}. In our models, NS remnants, when formed, are assigned randomly sampled magnetic fields (range $10^{11.5}-10^{13.8}\,\rm{G}$) and spin periods (range $30-1000\,\rm{ms}$) to match observations of young pulsars. To model NS evolution, we follow the prescriptions described in \cite{hurley2002evolution} and \cite{Kiel_2008}.

As outlined in \citet{Kiel_2008}, we assume that the dominant spin-period evolution mechanism for single NSs is dipole radiation, and NSs are treated as solid spheres. The spin-down rate of single NSs is \begin{equation} \dot{P}=K\frac{B^2}{P},\end{equation} where $K=9.87\times10^{-48}\,\rm{yr/G^2}$, $B$ is the surface magnetic field and $P$ is the spin period\footnote{Note that there is a typo in the value of $K$ in \cite{Kiel_2008}; our adopted value is the correct one.}. 

Additionally, magnetic fields of single NSs are assumed to decay exponentially, \begin{equation} B=B_0\exp\left(-\frac{T}{\tau}\right), \end{equation} on a timescale $\tau=3$ Gyr \citep{Kiel_2008}. Here $B_0$, and $T$ are the initial magnetic field, and the NS's age, respectively. \citet{faucher2006birth} showed that there is no significant magnetic field decay for single radio pulsars on a timescale of $\sim 100\,\rm{Myr}$. Recent magneto-thermal models of NS magnetic field also show that the magnetic field evolution of single radio pulsars is compatible with no decay or weak decay \citep[e.g.,][]{popov2010population,vigano2013unifying}. The timescale we adopt here for magnetic field evolution of single radio pulsars is compatible with a very weak field decay. Further exploration of the effects of magnetic field decay on GC pulsars will be presented in future works.

For NSs in binaries, binary evolution is also taken into account. The evolution of NSs in detached binaries is the same as for single NSs. On the other hand, during mass-transfer episodes, the magnetic fields and spin periods of NSs can change significantly on a short timescale. During accretion, the magnetic field is assumed to decay as 
\begin{equation}
B=\frac{B_0}{1+\frac{\Delta M}{10^{-6} M_{\odot}}}\exp(-\frac{T-t_{\rm{acc}}}{\tau}),
\end{equation} where $t_{acc}$ is the duration of the NS accretion phase (during RLOF) and $\Delta M$ is the mass accreted\footnote{We use the standard value $10^{-6}\,M_{\odot}$ for the threshold accreted mass, as adopted in \texttt{COSMIC}. This ensures that the first factor acts as a simple switch, lowering the NS magnetic field as soon as even a small amount of mass is accreted. Changing the value adopted here, as long as it is very small, produces negligible changes in the overall results.}. The NS spin period decreases accordingly through angular momentum transfer \citep[see][equation~(54)]{hurley2002evolution}. Wind mass loss, tidal evolution, magnetic braking and supernova kicks have only small effects on the magnetic field and spin period evolution. Only stable mass transfer in a binary system can spin up NSs to MSPs in our models; we ignore the possibility of mass accretion by NSs during common-envelope phases \citep{hurley2002evolution}.

Occasionally, through dynamical or binary evolution, NSs merge with main-sequence (MS) stars, giants, or WDs. If the outcome of this merger is a NS (as opposed to a BH; \citealp[see][Table~2]{hurley2002evolution}), the magnetic field and spin period for this NS are reset by drawing from the same ranges of initial values as above. However, if a MSP is involved in such a merger, different initial magnetic fields and spin periods are assigned (relative to regular NSs) so that the new values still match those observed in MSPs. Specifically, for MSPs involved in mergers, a new magnetic field is drawn from the range $10^{8}-10^{8.8}\,\rm{G}$ and a new spin period is drawn from the range $3-20\,\rm{ms}$. In other words, we assume that a MSP involved in a merger remains a MSP after the merger, but allowing for a small random change in magnetic field and spin period.

In addition, for all MSPs, we assume a lower limit for the NS magnetic fields to be $5 \times 10^7\,\rm{G}$ \citep{Kiel_2008}. We do not set a lower limit for the spin periods. We use the standard expression
\begin{equation}
\frac{B}{P^2}=0.17\times10^{12}\,\rm{G\ s^{-2}}
\end{equation} 
\citep{ruderman1975theory,bhattacharya1992decay} for the pulsar death line. Note that the exact location of the death line is still under debate \citep[see, e.g.,][]{zhang2000radio,zhang2002radio,zhou2017dependence}. 
Throughout this paper we define a pulsar as any NS with magnetic field and spin period above the death line, a MSP as any pulsar with $P \lesssim 30\,$ms, and a young pulsar as any pulsar that is not a MSP.

\subsubsection{Electron-capture Supernovae}\label{subsubsec:ECSN}
In our models, NSs formed in ECSNe are the dominant type of retained NSs in the cluster, and in all NS-LMXBs and MSPs. We assume that an ECSN happens whenever an ONeMg WD reaches a critical mass $M=1.38\,M_{\odot}$ so that electron capture is triggered on $\rm{Mg}^{24}$ and $\rm{Ne}^{20}$ and the WD undergoes collapse from the sudden lack of electron pressure support \citep{miyaji1980supernova,nomoto1984evolution,nomoto1987evolution}. We give small natal kicks to ECSN NSs, drawn from a Maxwellian distribution with a dispersion $\sigma_{ECSN}=20\,\rm{km\,s^{-1}}$ \citep{Kiel_2008}.

In \texttt{CMC} we have three different evolutionary paths that lead to ECSNe \citep{Ivanova_2008}. The first one is evolution-induced collapse (EIC) of a single star with initial mass in the range $6-8\,M_{\odot}$ \citep{nomoto1984evolution,nomoto1987evolution,Kiel_2008}. For main-sequence (MS) and giant stars, if their initial masses are smaller than the carbon ignition mass and larger than the critical NS formation mass, they can form NSs in ECSNe. Helium stars with masses between $1.6\,M_{\odot}$ and $2.25\,M_{\odot}$ \citep{hurley2000comprehensive} also go through ECSNe. The second path is accretion-induced collapse (AIC) \citep{nomoto1991conditions,saio2004off}. If an ONeMg WD accretes ONe or CO material from its companion during RLOF, the WD is assumed to collapse to a NS when its mass is larger than the ECSN critical mass and smaller than the maximum NS mass set by \texttt{BSE}, above which it will become a BH. The last path is the merger-induced collapse (MIC) of the product of a merger or collision between two WDs \citep{1985A&A...150L..21S}; these can be either ONe WDs or CO WDs. All three paths generally produce NSs in binaries, which can often lead to subsequent RLOF and the production of LMXBs and MSPs.

It is important to note that the mass range for ECSNe progenitors is under debate \citep[see, e.g.,][ and references therein]{poelarends2017electron}. However, changing the mass range for EIC in our models has only a small effect on the number of retained NSs and MSPs at late times. This is because, by assumption, the ECSN kicks are not mass dependent. In general, since the mass range for helium cores to collapse in ECSNe is narrow, small changes to the EIC mass range will not affect the overall number of EIC NSs and MSPs significantly. Thus we simply adopt the \texttt{SSE} prescription as described above.

\subsection{Influence of Dynamics on MSP Formation}\label{subsec:dynamics}
After their formation at early times, most of the NSs in our models evolve like young pulsars in isolation until the end ($12\,\rm{Gyr}$). Their magnetic fields and spin rates slowly decrease until they die as pulsars. Dynamical interactions are essential in producing MSPs throughout the evolutionary histories of GCs \citep[e.g.,][]{Ivanova_2008}. Mass segregation puts NSs close to the cluster centers in high stellar density regions. During frequent stellar encounters, single NSs can acquire companions through exchanges, and wide binaries can be hardened by repeated interactions. NSs in binaries can then be spun up to MSPs through mass transfer.

\begin{figure}
\begin{center}
\includegraphics[width=\columnwidth]{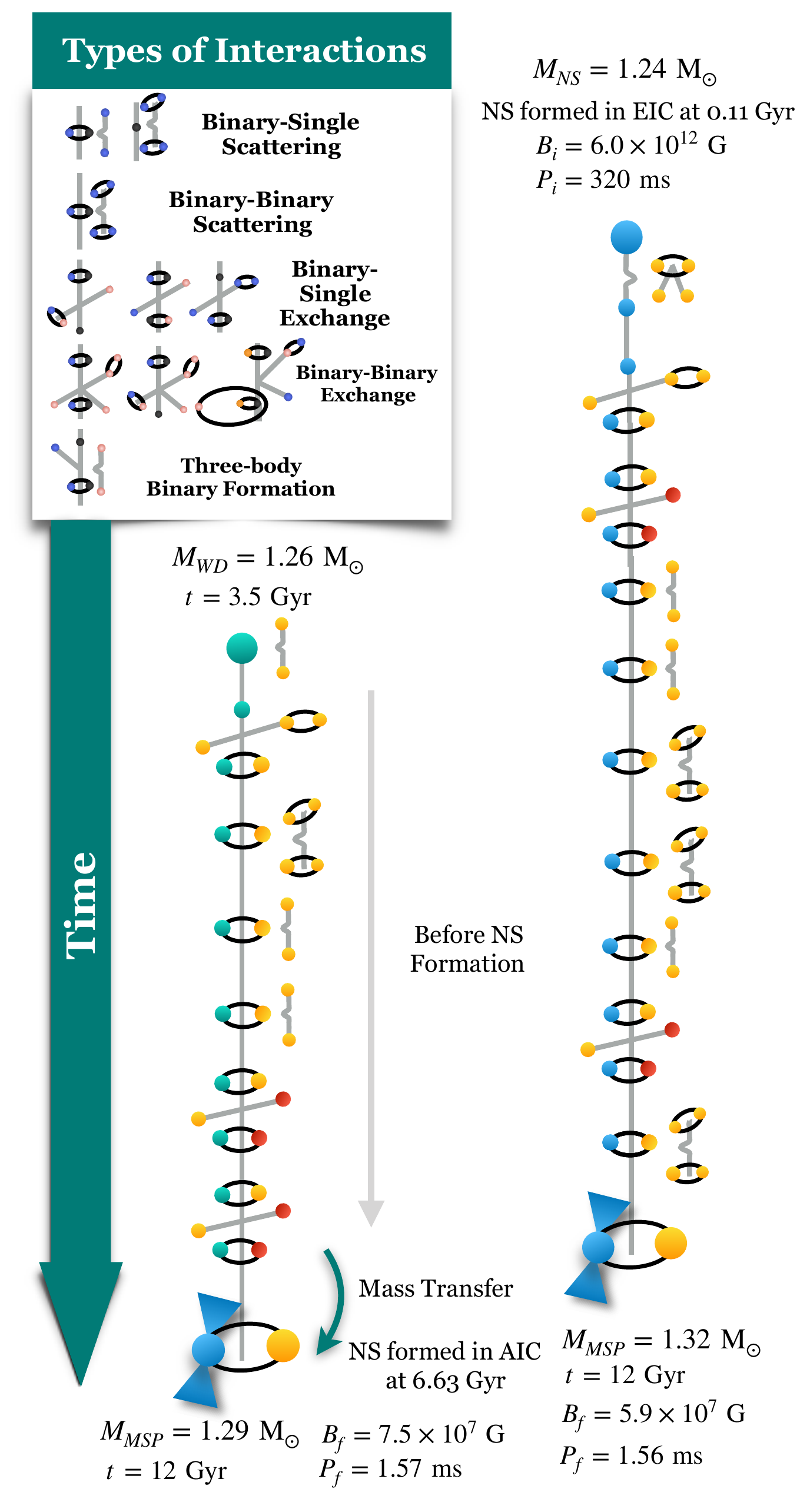}
\caption{Illustration of dynamical interactions of two NSs that become MSPs at late times. On the right, blue dots denote the NS that is spun up to a MSP. On the left, green dots denote the WD that collapses to a NS and becomes a MSP at late times. Other stars that are either their binary companions or dynamically interact with the NSs are shown by yellow and red dots. \label{fig:dyndiagram}}
\end{center}
\end{figure}

\begin{figure}
\begin{center}
\includegraphics[width=\columnwidth]{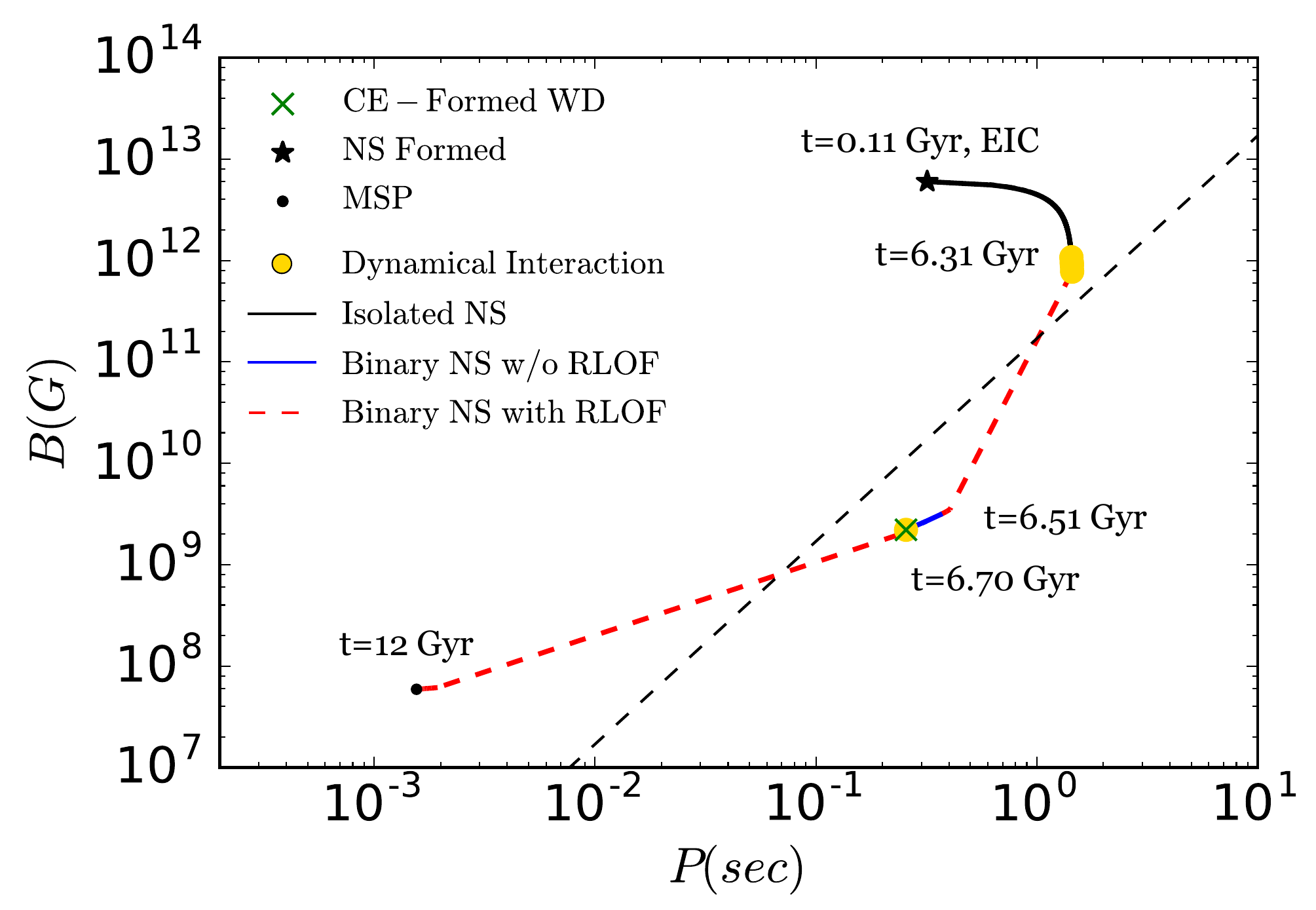}
\caption{Magnetic field and spin-period evolution of a NS that becomes a MSP at late times. This NS was formed from EIC. The black star and the black dot mark the progenitor NS formation time and the MSP at $12\,\rm{Gyr}$, respectively. The yellow dots show different dynamical interactions. The green cross marks the time when the companion star became a WD following a common envelope phase. The black, blue and red paths indicate the NS is in isolation, in a detached binary, or going through mass accretion, respectively. The black dashed line shows the death line (below which the radio pulsar emission is assumed to turn off). Clearly dynamical interactions play the key role in the formation of this typical MSP.\label{fig:EIC_BP}}
\end{center}
\end{figure}

Figure~\ref{fig:dyndiagram} illustrates two examples of the dynamical interaction histories of MSPs. The progenitor NS of a MSP on the right of the figure was formed in EIC at $0.11\,\rm{Gyr}$. It had a magnetic field of $6\times10^{12}\,\rm{G}$ and a spin period of $320\,\rm{ms}$ at birth. The pulsar experienced its first encounter at $5.3\,\rm{Gyr}$ and acquired a WD companion through a binary-single exchange. A second binary-single exchange replaced the WD companion to another WD. The binary then experienced a few binary-single and binary-binary scatterings. During this time, there was no mass transfer, and the magnetic field and spin rates of the pulsar were decreasing (solid black line in Fig.~\ref{fig:EIC_BP}). A third binary-single exchange encounter gave the pulsar a MS star companion. The MS star spun up the pulsar via RLOF-driven mass transfer (the first red-dashed line in Fig.~\ref{fig:EIC_BP}). At this time it was not yet a MSP (not all mass transfer leads to MSPs, only extended periods of stable mass transfer can produce MSPs). The MS star later evolved to become a giant and the binary underwent a common-envelope phase (blue line in Fig.~\ref{fig:EIC_BP}), which circularized and shrank the binary orbit. The remnant of the MS star was a WD which continued to fill its Roche lobe and spun up the pulsar to a MSP (The second red-dashed line in Fig.~\ref{fig:EIC_BP}). The final system has a MSP with a $5.9\times10^7\,\rm{G}$ magnetic field and a $1.57\,\rm{ms}$ spin period, and a companion with mass $0.0075\,M_{\odot}$ at $12\,\rm{Gyr}$. Note that the companion mass is very small because it has been depleted through the extended period of mass transfer that spun up its pulsar companion, as in ``black widow" type binary MSPs \citep[e.g.,][]{rasio2000formation}.

The MSP on the left of the figure has a different evolutionary path. The NS was not formed until about $6.6\,\rm{Gyr}$ in AIC as a member of a double WD binary, which itself was formed through a series of dynamical interactions, including both binary-single scatterings and exchanges. After the formation of the NS there were no stellar encounters and the WD companion kept transferring mass to spin up the NS to a MSP. The MSP at $12\,\rm{Gyr}$ has a magnetic field of $7.5\times10^7\,\rm{G}$ and a spin period of $1.56\,\rm{ms}$, again with a low-mass companion of mass about $0.0075\,M_{\odot}$.

\begin{figure}
\begin{center}
\includegraphics[width=\columnwidth]{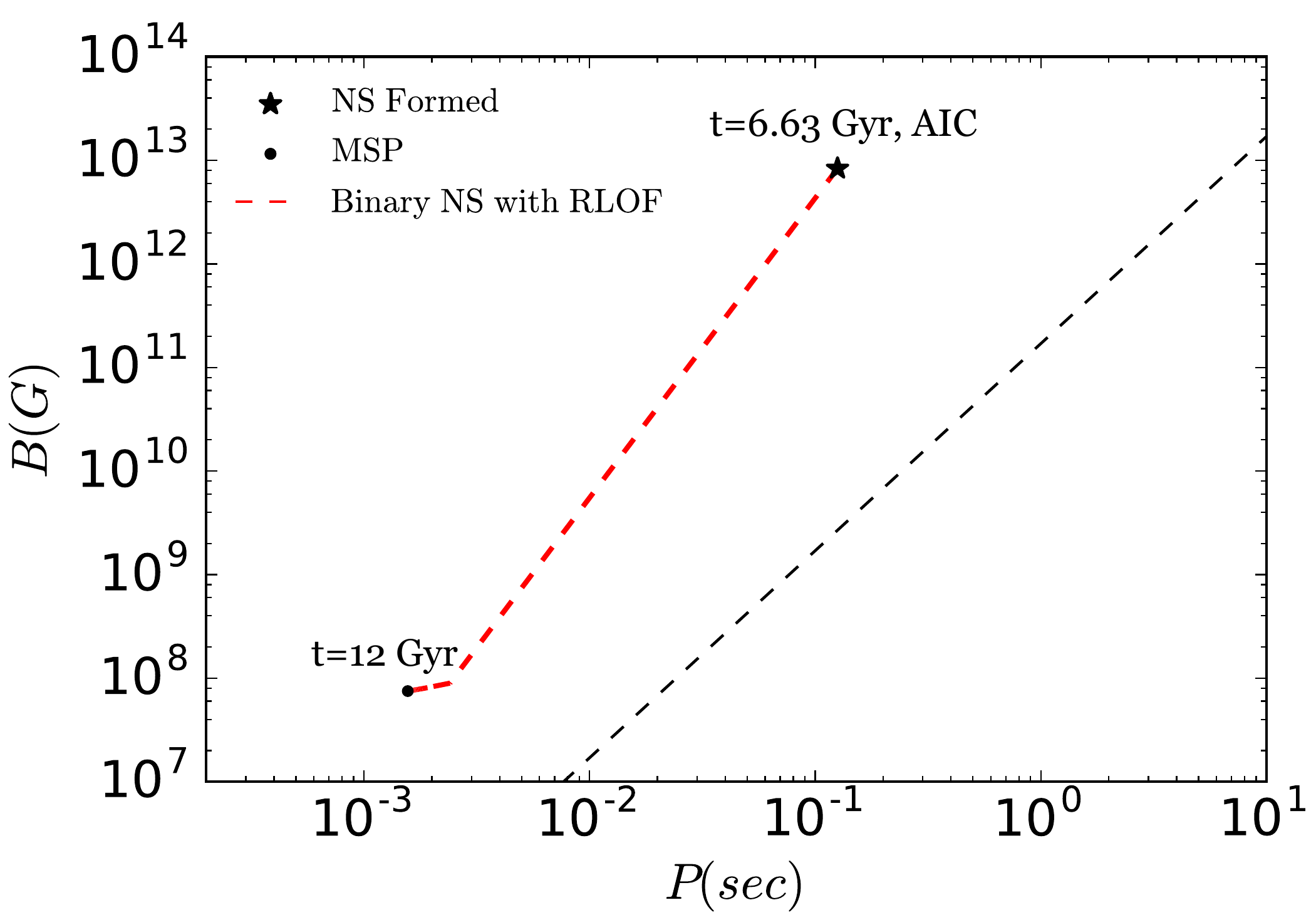}
\caption{Same as Figure~\ref{fig:EIC_BP} but for a progenitor NS formed through AIC. Note that, although there is no dynamical interaction shown here, the AIC NS is a product of multiple interactions, as shown in Figure~\ref{fig:dyndiagram}. \label{fig:AIC_BP}}
\end{center}
\end{figure}

Figures~\ref{fig:EIC_BP} and~\ref{fig:AIC_BP} show the evolution of magnetic fields and spin periods of the same two MSPs as in Figure~\ref{fig:dyndiagram}. The black dashed line is the death line (see equation (4)), below which radio pulsars disappear as they can no longer support pair production. 

In Figure~\ref{fig:EIC_BP}, for the first 6~Gyr the magnetic field of the pulsar decreased due to magnetic field decay, and the spin period increased slowly due to magnetic dipole radiation. The pulsar exchanged into a binary at $6.31\,\rm{Gyr}$ with a MS star companion, and was accreting material from the MS star for about 200 Myr, causing it to spin up from angular momentum transfer. During this time the pulsar temporarily went below the death line. The MS star then evolved to a giant star and went through a common envelope phase, during which the pulsar was not accreting mass. The binary orbit was circularized and shrunk by the common envelope. At about $6.7\,\rm{Gyr}$, the MS star became a WD and started mass transferring, continuing to spin up the pulsar. The pulsar emerged out of the ``graveyard" and was spun-up to a MSP at late times.

Figure~\ref{fig:AIC_BP} is similar to Figure~\ref{fig:EIC_BP} but for the other MSP on the left side of Figure~\ref{fig:dyndiagram}. There were no dynamical interactions between the time the NS formed and $12\,\rm{Gyr}$. In this case the NS-WD binary stayed intact after AIC and the WD continued through RLOF and spun the pulsar up to a MSP. A few studies have suggested that NSs formed in AIC are unlikely to be spun up to MSPs in the same binary  \citep[e.g.,][]{Verbunt1987xraybinary}, because MSPs formed in this way cannot explain the overabundance of LMXBs and MSPs in GCs. However, this is in contrast to the results from our models. In the progenitor binaries of AIC NS, the companions may fill the Roche-lobe and transfer mass onto the ONe WD, which likely leads to common envelope evolution in \texttt{BSE}. The common envelope evolution circularizes and shrinks the binary orbits, producing tight (semi-major axis about $0.001\,\rm{AU}$) WD--WD or NS--WD binaries. Because the binaries are so tight, the same WD companions can continue to fill the Roche lobe and spin up the NSs to MSPs.

\section{RESULTS} \label{sec:results}
\subsection{Overview of Statistics}
Consider all NSs still in our model clusters at 12 Gyr. These represent about $8\%$ of all NSs ever born in the $N=8 \times 10^5$ models, and about $15\%$ in the $N=3.5 \times 10^6$ model. Of all these retained NSs at 12 Gyr, in the $N=8 \times 10^5$ models (models 1-25), about $5\%$ came from CCSN NSs, while about $95\%$ came from ECSNe. For the large-$N$ model (model 26), about $15\%$ NSs came from CCSNe, and about $85\%$ came from ECSNe. For models with fewer retained BHs, a larger fraction of NSs retained at 12 Gyr were formed through dynamically influenced channels (mainly AIC and MIC): about $0.5\%$ of the NSs that are retained in BH-rich (models 1-10) and BH-intermediate models (models 11-19) and about $5\%$ in BH-poor models (models 20-26) are formed through these channels. In all the models, most of the retained NSs at 12 Gyr were formed through EIC.

Now we turn to the sample of NSs in our models that would be potentially detectable as radio pulsars at 12 Gyr (all NSs above the death line defined by equation~(4)). In total, there are $208$ pulsars in all our models, including $142$ MSPs (i.e., pulsars with $P\lesssim30\,$ms) and $66$ young pulsars. Note that these numbers do not take into account beaming (and the beaming fraction is thought to be much smaller for young pulsars; see \citealp[e.g.][and references therein]{lorimer2008binary}). Because CCSN NSs are such a small fraction of all NSs retained at late times, this class of NS does not contribute significantly to the pulsar population observable in old GCs. In contrast, $26\%$ of all MSPs in our models were formed through EIC, $68\%$ through AIC, and only $5\%$ through CCSNe and $1\%$ through MIC. Therefore ECSN NSs are the principal source for MSPs in GCs. MIC NSs does not contribute much in general to MSPs. This is because there are only a few MIC NSs in the models (the total number is comparable to CCSN NSs in BH-poor models, but almost zero in the other models), and most of them formed at late times (they do not have enough time to be spun up).

Most of the MSPs in our models are in binaries, with a binary fraction of about $80\%$, and most of the companions are low-mass stars (only 9 companions are MS stars). There are also a few DNSs and NS--BH binaries in our models (see Sec.~\ref{subsec:DNS_NSBH}). Single MSPs result mostly from dynamical encounters in which the MSPs are exchanged out of the binaries.

\begin{figure}
\begin{center}
\includegraphics[width=\columnwidth]{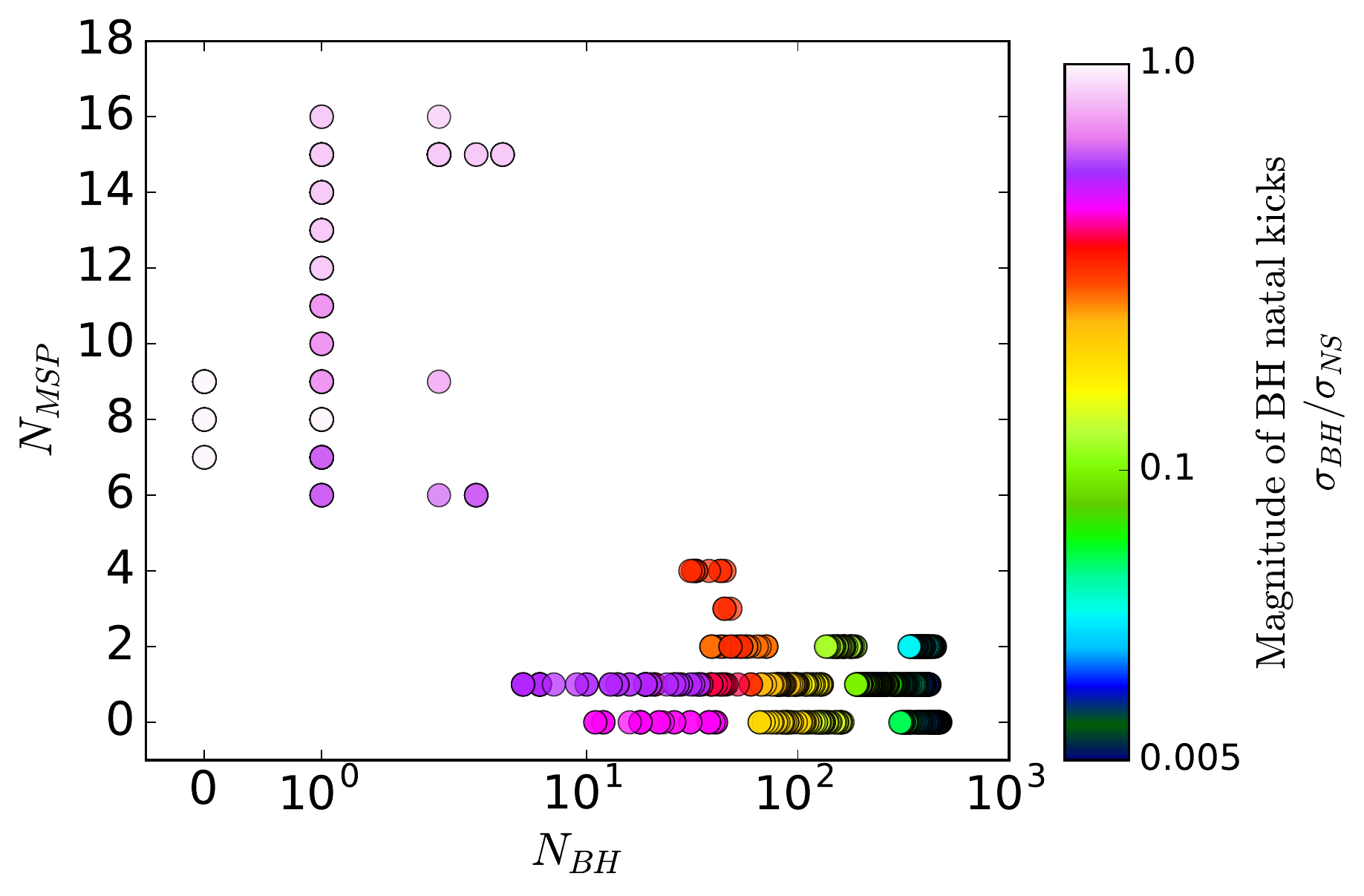}
\caption{Number of MSPs vs number of BHs between $9$ and $12\,\rm{Gyr}$ in models 1-25. Models with different kicks are shown with different colors. There is a clear anti-correlation between these two numbers.
\label{fig:nbh_nmsp}}
\end{center}
\end{figure}

\subsection{BH--MSP Anti-correlation}\label{subsec:BHMSP}
We find a clear anti-correlation between the number of retained BHs and the number of MSPs in our models. Figure~\ref{fig:nbh_nmsp} shows the number of BHs and MSPs between $9$ and $12\,\rm{Gyr}$ in models 1--25. We include multiple points from the same model sampled at different times, so a single model can provide different numbers of BHs and MSPs in the figure. For example, there are 4~models at $N_{BH}=1$, showing a range of $N_{MSP}$ the models can have for the same BH number. For cluster models with only a few retained BHs, there can be as many as 16~MSPs in the cluster; while for cluster models with more than 200~BHs, there are at most 2~MSPs. In the large-$N$ model (model~26) there are 83~MSPs and 20~BHs at $12\,\rm{Gyr}$, which we do not include in the figure, simply so that we can isolate the BH--MSP relation for models with similar $N$. It is worth noting that if the AIC MSPs are excluded from Figure~\ref{fig:nbh_nmsp}, the anti-correlation between the numbers of BHs and MSPs still holds.

\begin{figure}
\epsscale{2.4}
\plottwo{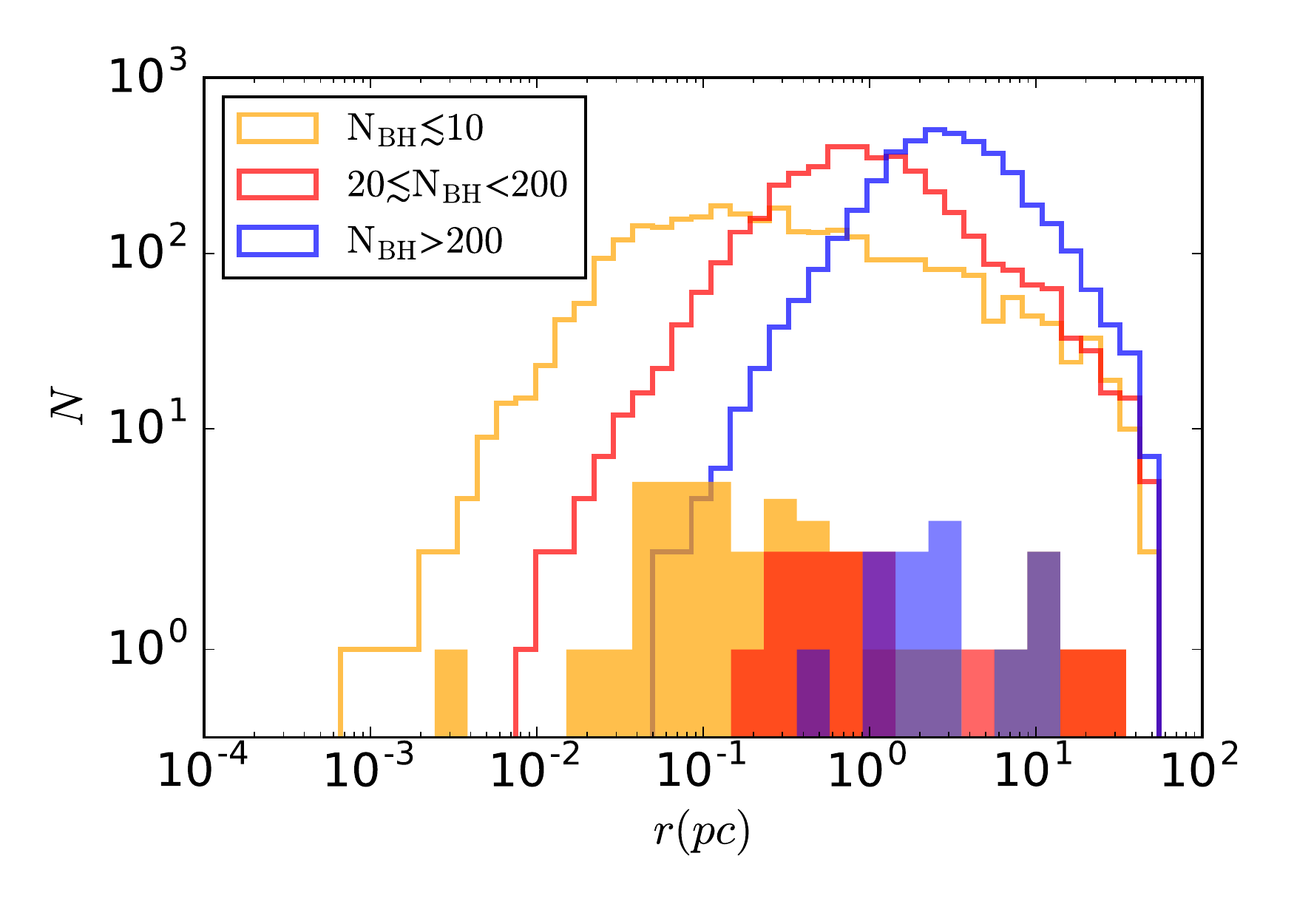}{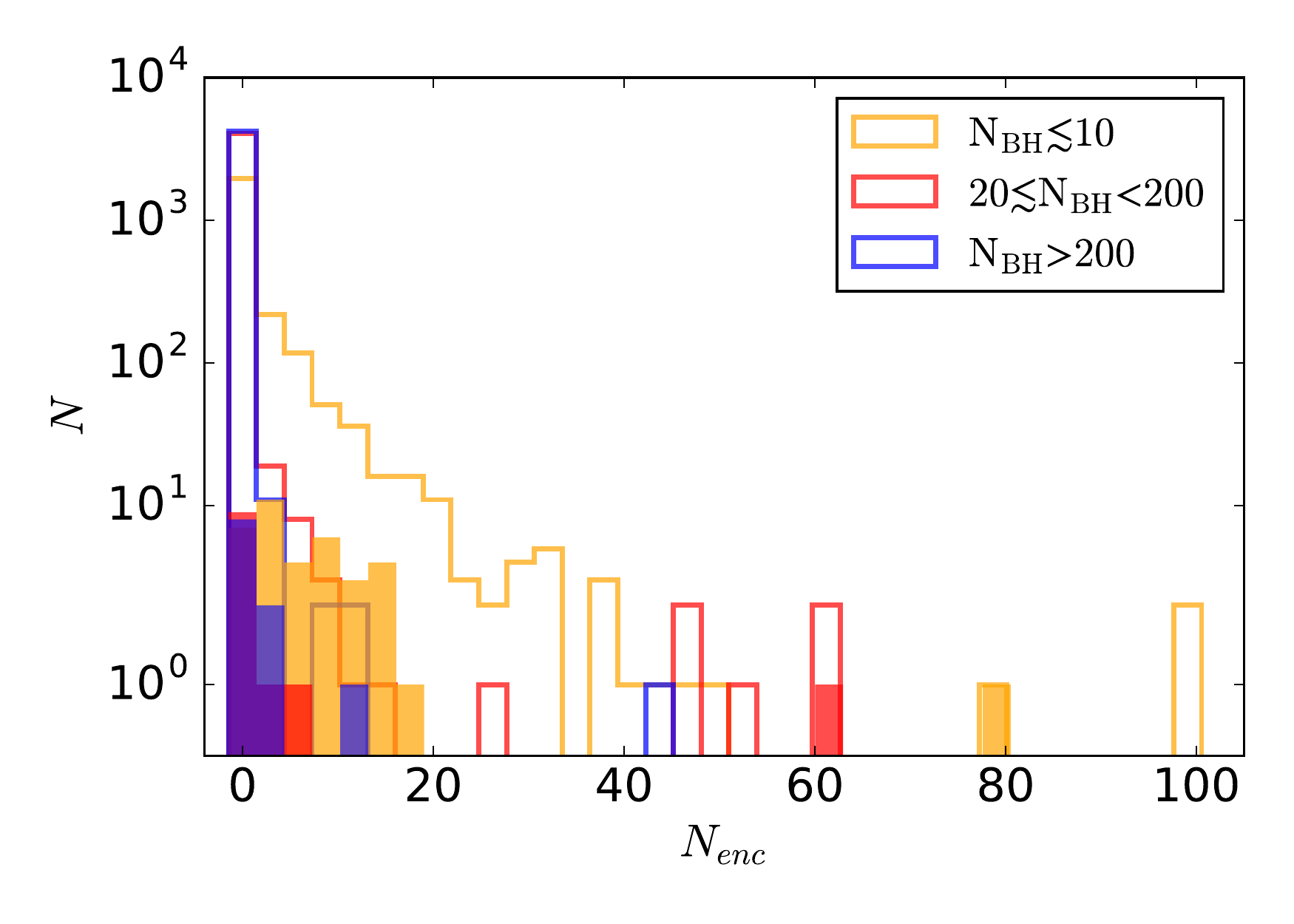}
\caption{Top: 2D-projected radial distribution of NSs in models 1-25. The yellow, red, and blue distributions correspond to BH-poor, BH-intermediate, and BH-rich models, respectively. The step histograms show the radial distribution of all the NSs in these models. The medians of the NS radial distribution are $0.25, 0.88, 2.86\,\rm{pc}$ for the three histograms. It is clear that, for models with a large number of BHs, the NSs are prevented from segregating close to the cluster centers; while for models with a small number of BHs, the NSs are much more centrally concentrated. The filled histograms show the radial distributions of the MSPs in the models. The MSP radial distributions show similar trends as the radial distributions for all NSs. Bottom: Number of encounters of NSs in models 1-25. Again, step histograms are for all the NSs, and filled histograms are for MSPs only. For models with a smaller number of BHs, the NSs experience more stellar encounters because they are closer to the center, where stellar densities are high. This can also be seen in the top figure, where the histograms for the BH-poor models are wider, showing that the NSs in these models interact more often. \label{fig:ns_radiusenc}}
\end{figure}

This anti-correlation was anticipated, based on our understanding of BH populations in GCs. Figure~\ref{fig:ns_radiusenc} shows how the BHs dynamically influence the NSs including the MSPs in the clusters. As long as many are present, BHs dominate the cluster cores because of mass segregation and they prevent the NSs from concentrating in the high-density central region. Furthermore, GCs with more retained BHs have lower core densities due to the heating of the cores from BH interactions \citep{fragione2018neutron,arca2018mocca}. The upper panel of Figure~\ref{fig:ns_radiusenc} shows the distribution of the 2D-projected radii of all NSs (step histograms) and only MSPs (filled histograms) in models 1-25. For models with fewer than 10 BHs, the NSs are located closer to the GC centers; while for models with a large number of BHs, the projected radii for most of the NSs are about an order of magnitude larger. This also affects the number of encounters the NSs can have during the cluster evolution. The stellar densities are higher towards the cluster centers, where the average stellar densities within the median 2D-projected radii (Fig.~\ref{fig:ns_radiusenc}) for the BH-rich, BH-intermediate and BH-poor models are about $6.6\times10^3\,\rm{pc^{-3}}$, about $8.9\times10^4\,\rm{pc^{-3}}$ and about $1.0\times10^6\,\rm{pc^{-3}}$, respectively. As a result, NSs located closer to the centers go through more dynamical interactions. Therefore NSs in the BH-poor models are more likely to acquire companions and accrete mass, and thus, a larger chance to become MSPs at late times. The numbers of encounters of all NSs and MSPs are shown in the lower panel of Figure~\ref{fig:ns_radiusenc}. This trend can also be seen in the upper panel, where NSs in the BH-poor models are scattered more and have a wider radial distribution, in contrast to NSs in the BH-rich models.

The number of BHs can also be constrained by other measurable quantities such as the core radius of a cluster, which is strongly correlated with the number of retained BHs (Table~\ref{tab:table_1}). In our models, retaining more BHs leads to larger core radii, lower central (non-BH) stellar densities, and lower rates for all dynamical interactions. Thus the basic properties of the MSP population (number and radial profile) and the core radius correlate similarly with the number of BHs present in a cluster. There might also be intrinsic differences between the properties of the MSPs in denser versus less dense clusters (such as systematic differences in spin periods or binary companion types). However, we do not have sufficient coverage of the statistics in either the observed sample (also affected strongly by selection bias) or in our current set of models (too few MSPs in low density clusters; 11 total in models 1-10) to study this at present.

Metallicity has a minor effect on the number of retained NSs in clusters and is unlikely to affect this anti-correlation between the number of retained BHs and the number of MSPs in a cluster. However, we intend to perform a more detailed study of the effects of changing metallicity in these models in future work.

\begin{figure}
\begin{center}
\includegraphics[width=\columnwidth]{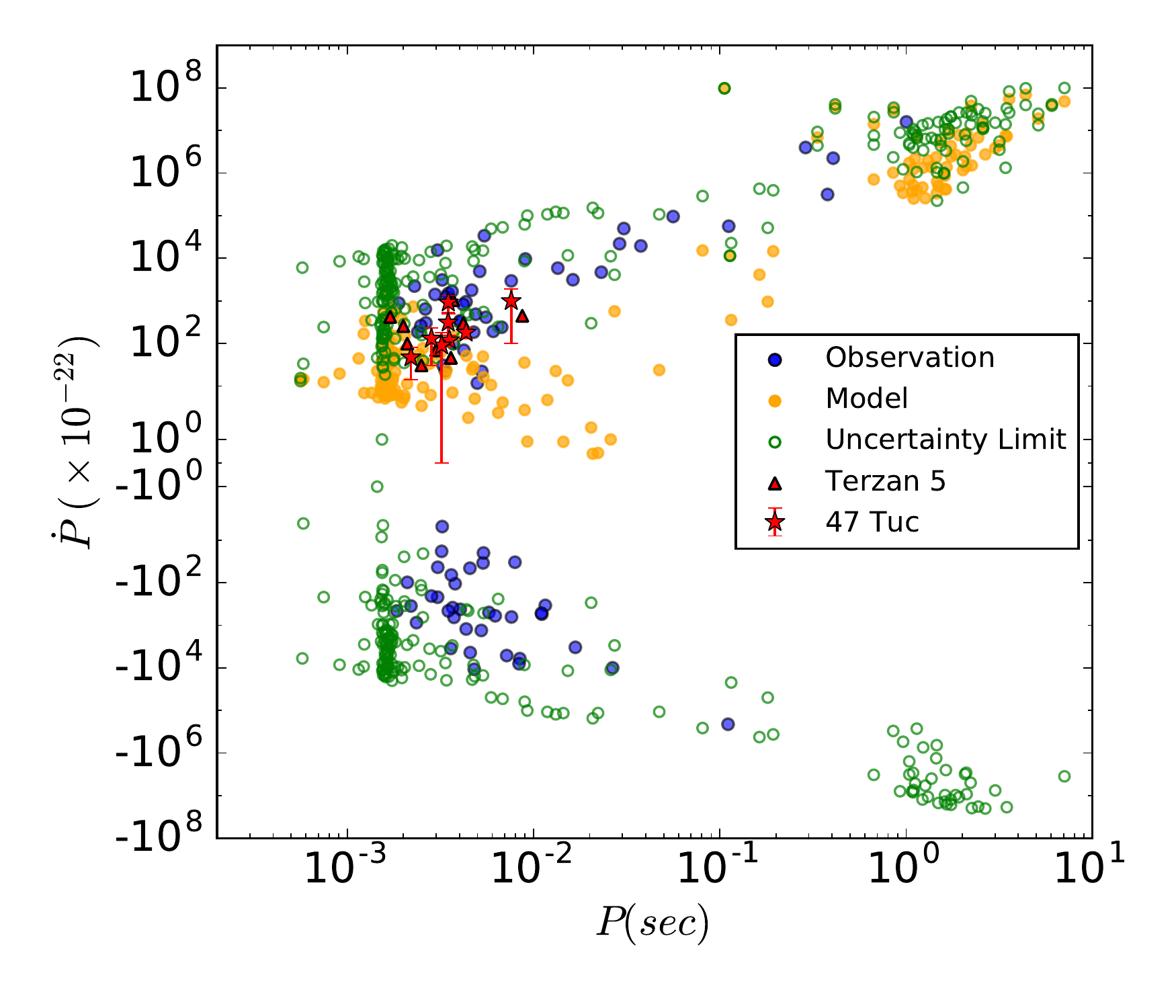}
\caption{Spin periods and spin period derivatives of pulsars. The blue dots show all observed GC pulsars. The red stars and triangles show the pulsars in 47~Tuc with derived intrinsic spin period derivatives and in Terzan~5 with inferred magnetic fields. The orange dots show the intrinsic $\dot{P}$ values for our model pulsars; the green dots show the corresponding apparent $\dot{P}$ values taking into account acceleration in the cluster potential. \label{fig:ppdot}}
\end{center}
\end{figure}

\subsection{Spin Periods and Spin Period Derivatives}\label{subsec:PPdot}
We plot all the pulsars in our models on top of all observed pulsars in the GC pulsar catalog\footnote{\url{http://www.naic.edu/~pfreire/GCpsr.html}} in Figure~\ref{fig:ppdot}. The intrinsic spin period derivatives of the model pulsars are derived from their magnetic fields using $\dot{P}=K\frac{B^2}{P}$ (see Sec.~\ref{subsubsec:BP}). In the figure, the blue dots show the spin periods and spin period derivatives of the observed GC pulsars and the orange dots show the intrinsic $P$ and $\dot{P}$ values for model pulsars. We calculate the maximum accelerations the pulsars can have in their cluster potential using Eq.~(2.5) of \cite{phinney1992pulsars}. The upper and lower limits for the``observable" $\dot{P}$ values of model pulsars with accelerations are shown as green circles. We also include a few pulsars in 47~Tuc with derived intrinsic $\dot{P}$ values, and in Terzan~5 with inferred magnetic fields, in Figure~\ref{fig:ppdot}; these are shown by red stars and triangles, respectively \citep{freire2017long,prager2017using}.

There are both MSPs ($P \lesssim 30\,$ms) and young pulsars in our models (Fig.~\ref{fig:ppdot}), as are observed in GCs. Most of the pulsars observed in Galactic GCs are MSPs, as expected since MSPs have long lifetimes and can exist in GCs for many Gyr. In contrast, young pulsars have relatively short lifetimes, and those formed at early times in GCs are no longer there. However, through dynamical interactions such as collisions between a MS star and a WD, young pulsars can be formed at the present time in old GCs. Almost all young pulsars in our models were formed by collisions at late times (Fig.~\ref{fig:ppdot}), including newborn NSs formed in WD--MS star or WD--WD collisions, and old NSs that partially accreted during collisions with MS or giant stars. It has also been suggested that (apparently) young pulsars in GCs could be formed by partial recycling of NSs in mass-transferring binaries \citep{verbunt2014disruption}. However, this does not happen at a significant rate in our models: only 3~young pulsars in our models were formed by partially spinning up a NS through accretion in a binary. 

Figure~\ref{fig:ppdot} shows that the spin periods and spin period derivatives of our model pulsars are in reasonable agreement with observations, suggesting that our current very simple treatment for the evolution of magnetic fields and spin periods of NSs in {\tt CMC} is sufficient for this first attempt at a detailed comparison.

\subsection{DNSs and NS--BH Binaries}\label{subsec:DNS_NSBH}
DNSs and NS--BH binaries provide important probes for general relativity and their mergers can now be detected as gravitational wave sources and can power short gamma-ray bursts visible throughout the universe \citep[e.g.,][]{belczynski2002study,clausen2014dynamically}. Such systems may form in a variety of astrophysical environments, including through isolated massive binary evolution in  galactic fields \citep[e.g.,][]{belczynski2002comprehensive}, as well as through dynamical processes in dense star clusters such as galactic nuclei \citep[e.g.,][]{fragione2018black} and GCs \citep[e.g.,][]{Sigurdsson1995binaryandneutronstar}.

So far only one DNS system has been identified in a GC, PSR 2127+11~C in M15\footnote{There is also a candidate DNS in NGC~6544; see \citet{lynch2012timing}.}. M15C has a spin period of $30.5\,\rm{ms}$ and a derived magnetic field of $1.2\times10^{10}\,\rm{G}$. Both the pulsar and its NS companion have masses $1.35\,M_{\odot}$. This DNS system has an 8-hour orbital period and a highly eccentric orbit with $e=0.68$ \citep{anderson1990discovery,prince1991timing,deich1996massm15c,jacoby2006measurement}. The projected radius of M15C from the center of M15 is $2.7\,\rm{pc}$, which, combined with its eccentricity, suggests that it was most likely formed by an exchange interaction in the cluster core with recoil to its current location \citep{Phinney:1991aa}.

Primordial DNSs must be very rare in GCs because of natal kicks for NSs that break up the progenitor binaries or eject them out of the clusters. However, one way of producing these systems in GCs is through dynamical interactions. M15 is ``core-collapsed", i.e., it has an extremely high stellar density near its center \citep[][2010 edition]{harris1996catalog}, which provides a good environment for DNS formation. NS--BH binaries should be even more rare than DNSs in GCs, even taking dynamics into account. Indeed, in clusters with many BHs, mass segregation and the strong heating by BH interactions prevent the NSs from interacting with the BHs (see Fig.~\ref{fig:ns_radiusenc}), while in clusters with few BHs, NSs have few interactions with BHs compared to other stars. It is therefore not surprising that no NS--BH binary has ever been detected. Based on simple models of the dynamical evolution of BH binaries in fixed GC backgrounds (described by multi-mass King models), \cite{clausen2014dynamically} estimate that there are at most $\sim 10$ BH--MSP binaries in the entire Milky Way GC system. We plan to further explore the formation and merger rates of NS--BH binaries in GCs in future studies using {\tt CMC}.

In our current models we found that several NS--BH binaries and DNSs formed dynamically. The last two columns of Table~\ref{tab:table_1} show the total numbers of these systems that appeared between $9$ and $12\,\rm{Gyr}$ in our models. Two of the DNSs contain MSPs, while 32 DNSs and 6 NS--BH binaries contain young pulsars. Most of the NS--BH binaries and DNSs in our models appeared for a short time ($< 100\,\rm{Myr}$) and were subsequently disrupted by dynamical encounters. However, there are also a few that survived for longer times. For example, in the large-N model there is a long-surviving DNS formed through a binary--binary  interaction involving a MSP of $1.32\,M_{\odot}$ and a NS of $1.24\,M_{\odot}$ at about 11.5 Gyr. In the final model at 12 Gyr the binary has nearly completed its inspiral, reaching an orbital period of just $8\,\rm{min}$ (the remaining time to merger is about 0.03 Myr). 

In total, we identify $40$ unique NS--BH binaries and $93$ unique DNSs in our models. Of these, 11 NS--BH binaries and 22 DNSs have gravitational-radiation inspiral times less than a Hubble time, and 9 NS--BH binaries and 15 DNSs merged during the model clusters' evolution time. If such binaries merge in the local universe, they may be observed as gravitational-wave sources by detectors such as LIGO/Virgo \citep{abbott2017gw170817} and they may produce short-hard gamma-ray bursts \citep{meszaros2006gamma,lee2007progenitors}. A more detailed study of the merger rates and properties of NS--BH binaries and DNSs in GCs will be presented in a follow-up paper.

\section{Discussion and Summary}\label{sec:sum_discuss}
It is well understood that pulsar observations are limited by strong selection effects. The large distances of most GCs from Earth, Doppler shifts in short-period binaries, and dispersion of signals by the interstellar medium all make it difficult to detect radio pulsars. The luminosity function of GC pulsars, $d\log\, N \approx -d\log\, L$ \citep{hessels20071,hui2010dynamical}, suggests that some low-luminosity pulsars may remain unobserved.

Given these observational biases, the true number of pulsars in any given cluster remains quite uncertain. Many studies have attempted to constrain the true population of radio pulsars in Milky Way GCs \citep{kulkarni1990pulsar,hui2010dynamical,bagchi2011luminosities,chennamangalam2013constraining,turk2013empirical}. For example, the predicted numbers of potentially observable pulsars in Terzan~5 and 47~Tuc are about $150$ and $80$, respectively, with large error bars ($\sim50\%$) \citep{bagchi2011luminosities,chennamangalam2013constraining}. Assuming the beaming fraction to be about $50\%$, there are around $70$ potentially observable pulsars in our model~26 (Table~\ref{tab:table_1}), a very promising result since this model has a mass at 12~Gyr close to that of 47~Tuc \citep[e.g.,][]{giersz2011monte}.

There are $5$ conspicuously {\em sub-millisecond} pulsars (sub-MSPs) in our models, as shown in Figure~\ref{fig:ppdot}. The progenitor NSs were exchanged into binaries with eccentric orbits and with MS star or giant star companions in dynamical interactions, and they were spun up all the way to these very short spin periods through mass transfer during the evolution of their companion stars. However, there is no sub-MSP observed in any GC or in the Galactic field. Since we do not assume a lower limit for the NS spin period, sub-MSPs can appear in our models through continued mass transfer. The absence of detected sub-MSPs indicates that some physical mechanism not included in our models, such as gravitational radiation \citep[see e.g.][]{bildsten1998gravitational,chakrabarty2003nuclear}, may set a fundamental limit on how fast a NS can be spinning. Additionally, many of our model MSPs are still accreting (at very low mass-transfer rates; $\dot{M} \lesssim 10^{-11}\,M_{\odot} \rm{yr^{-1}}$) from extremely low-mass ($\lesssim 0.01\,M_{\odot}$) companions at late times. These binaries could potentially be observed as either radio pulsars or LMXBs. The companions may eventually be completely evaporated (as observed in ``black widow'' systems), a process that we do not currently include in our treatment of binary evolution. In the absence of a sufficiently detailed treatment of these systems in \texttt{BSE}, we have assumed all of them to be binary MSPs in our current models. 

About $20\%$ of all MSPs in our models are single. Of the remaining $\sim80\%$ found in binaries, about $75\%$ of MSPs have WD companions and about $5\%$ have MS star companions. Observationally, there are 17 black-widow systems and 12 red-back systems in the GC pulsar catalog. Assuming that the ratio between the numbers of black widows and red backs is not affected by selection effects, the fraction of MSP binaries with H-rich companions (red backs) in our models seems too low, and the fraction of MSP binaries with low-mass WD companions (black widows) seems too high, compared to observations. As this is our very first attempt at studying the formation of MSPs within our \texttt{CMC} code, for simplicity, we did not explore the various uncertainties associated with the treatment of binary evolution in \texttt{BSE}, and instead we focused on general trends (for example the anti-correlation between MSPs and BHs). We do get good overall agreement with observations, in the sense that our models produce reasonable numbers of all observed type of systems (single vs binary MSPs, very low-mass companions vs MS star companions, slow vs fast pulsars). More closely matching observations would require a more sophisticated treatment of the binary evolution physics, which is beyond the scope of this analysis, but will be considered in future works.

To summarize, in this study, we have simulated pulsars in GCs using our \texttt{CMC} code. We updated \texttt{BSE} to incorporate ECSNe for NS formation via three channels: EIC for single stars, and AIC and MIC for binary stars. For ECSN NSs, we apply low natal kicks with $\sigma_{ECSN}=20\,\rm{km\,s^{-1}}$. They are the major source for MSPs in our GC models. We also incorporate magnetic field and spin period evolution of NSs in \texttt{BSE} to model realistic pulsars. In our models, magnetic field evolution follows simple prescriptions for magnetic field decay or lowering the field through mass accretion. Spin periods then evolve according to the change in magnetic field and angular momentum of the NSs.

Dynamical interactions are clearly the key to understanding the formation of MSPs in GCs. Stellar dynamics can enhance the formation of NSs through AIC and MIC, which are likely to be retained in GCs, and enhance mass transfer in binaries with NS accretors. Furthermore, binaries can be harden or get to higher eccentricities through interactions, which leads to mass accretion and spinning up of NSs.

We have studied the pulsar population in 26 GC models. We found that the number of MSPs is anti-correlated with the number of BHs retained in GCs. This results from the dynamical coupling of BHs and NSs in our models, where most of the pulsars are formed dynamically, including the few young pulsars that are active at late times. Additionally, NS--BH and NS--NS binaries are also more readily formed in models with few BHs retained. The large specific abundances of MSPs in GCs relative to the Galactic field can be explained naturally by dynamical formation. This also provides a way to estimate the number of BHs in GCs given the number of observed pulsars. Furthermore, the spin periods and spin period derivatives of our model pulsars agree reasonably with observations, showing that \texttt{CMC} is able to model realistic pulsar populations in clusters. Our code can also produce a realistically large number of pulsars in a large-$N$ model, opening the door to future detailed modeling of very massive GCs such as 47~Tuc \citep{giersz2011monte}.

\acknowledgments
We thank Craig Heinke, Paulo Freire, Ron Taam, and the anonymous referee for useful discussions and comments on the manuscript. Our work was supported by NASA ATP Grant NNX14AP92G and NSF Grant AST-1716762 at Northwestern University. 
Computations were supported in part through the resources and staff contributions provided for the Quest high-performance computing facility at Northwestern. 
C.S.Y.\ acknowledges support from the NSF GK-12 Fellowship Program under Grant DGE-0948017. 
K.K.\ acknowledges support from the NSF Graduate Research Fellowship Program under Grant DGE-1324585.
S.C.\ acknowledges support from CIERA, and from NASA through Chandra Award TM5-16004X/NAS8-03060, issued by the Chandra X-ray Observatory Center
(operated by the Smithsonian Astrophysical Observatory for and on behalf of NASA under contract NAS8-03060).
C.L.R.\ is supported by a Pappalardo Postdoctoral Fellowship at MIT.
F.A.R.\ acknowledges support from NSF Grant PHY-1607611 while at the Aspen Center for Physics.
\software{\texttt{CMC} \citep{Joshi_2000,Joshi_2001,Fregeau_2003, fregeau2007monte, Chatterjee_2010,Chatterjee_2013b,Umbreit_2012,Morscher_2015,rodriguez2016million}, \texttt{Fewbody} \citep{fregeau2004stellar}, \texttt{BSE} \citep{hurley2002evolution}, \texttt{SSE} \citep{hurley2000comprehensive}}

\bibliographystyle{aasjournal}
\bibliography{MSP_BH_inGC_ApJ}

\end{document}